\documentclass[10pt,aps,prl,twocolumn,superscriptaddress]{revtex4-2}
\usepackage{graphicx}
\usepackage{tikz} 
\usepackage{amsmath}
\usepackage{braket}
\usepackage{comment}
\usepackage[export]{adjustbox}
\usepackage{qcircuit}
\usepackage{mathtools}
\usepackage{dblfnote}
\usepackage{subcaption}
\graphicspath{{images/}}
\usepackage{svg}
\usepackage{xcolor}

\usetikzlibrary{decorations.pathreplacing,calc}  

\begin{document}

%\begin{comment}

%\preprint{}

%\title{Enhancing Physics-Informed Neural Networks through the Integration of Variational Quantum Circuits}
\title{Quantum-Enhanced Convergence of Physics-Informed Neural Networks}


\author{Nils Klement}
%\thanks{}
\affiliation{Deutsches Zentrum f\"ur Luft- und Raumfahrt, Institut f\"ur Physik der Atmosph\"are, Oberpfaffenhofen, Germany}
\author{Veronika Eyring}
\affiliation{Deutsches Zentrum f\"ur Luft- und Raumfahrt, Institut f\"ur Physik der Atmosph\"are, Oberpfaffenhofen, Germany}
\affiliation{University of Bremen, Institute of Environmental Physics (IUP), Bremen, Germany}
\author{Mierk Schwabe}
\affiliation{Deutsches Zentrum f\"ur Luft- und Raumfahrt, Institut f\"ur Physik der Atmosph\"are, Oberpfaffenhofen, Germany}

\date{\today}

\begin{abstract}
Partial differential equations (PDEs) form the backbone of simulations of many natural phenomena, for example in climate modeling, materials science, and even financial markets. The application of physics-informed neural networks to accelerate the solution of PDEs is promising, but not competitive with numerical solvers yet. Here, we investigate how quantum computing can potentially improve the ability of physics-informed neural networks to solve partial differential equations. For this, we develop hybrid networks consisting of simulated quantum circuits combined with classical layers and systematically test them on various nonlinear PDEs and boundary conditions in comparison with purely classical networks. We demonstrate that the advantage of using quantum networks observed in state-vector simulations lies in their ability to achieve an accurate approximation of the solution in substantially fewer training epochs, particularly for more complex problems. These findings provide the basis for targeted developments of hybrid quantum neural networks with the goal to significantly accelerate numerical modeling. 
\end{abstract}
%\begin{comment}
%\keywords{}

\maketitle

%\section{Introduction}

\section{Introduction}
A key challenge in simulating real-world phenomena is the high computational demand to solve the underlying partial differential equations (PDEs). This leads not only to the need for high performance computational centers and large power consumption \cite{Suarez2025, fischer2020scalability}, but also limits the achievable resolutions. This is especially the case when real world data, such as observations or objectives, need to be included, and iterative numerical methods are therefore required. These requirements can be addressed in a unified manner by an approach called physics-informed neural network (PINN) \cite{issanayake1994}. For this purpose, not only the dynamics of the system, as described by a PDE, are incorporated into the loss function, but also arbitrary fixed points in space and time can be given, e.g. boundary conditions, observations, or objectives. An additional advantage of PINNs is that no discretization of space and time is needed, thus preventing discretization errors. Besides that, material properties \cite{wang2022rheology} or parametrized geometries \cite{10056991} can be treated as additional input variables, enabling the solution of a whole family of PDEs and the replacement of costly iterative methods. These features are especially relevant when the goal is to model real world phenomena supported by real world observations. For example, in climate models the fundamental dynamics are described by Navier-Stokes equations, while subgrid-scale processes are treated by parametrizations. These depend on free (tunable) parameters \cite{giorgetta2018icon}. To identify the optimal set of parameters, climate models must be calibrated against real-world observations \cite{Bonnet2025} which is commonly referred to as ``the inverse problem'' in the context of PINNs \cite{Kim2024}. Given the current urgency of enhancing climate models at present \cite{Schwabe2025a}, data-driven methods present promising advantages for this purpose. Another recent powerful application addresses the Navier–Stokes existence and smoothness \cite{PhysRevLett.130.244002, wang2025discoveryunstablesingularities}, one of the seven Clay Millennium Prize Problems. \\
Although in theory the possibilities and potential of classical PINNs (cPINNs) are promising, in reality they can not replace traditional numerical solvers yet. This is mostly due to trainability due to the complex loss - or in other words the computational costs - and a related lack of performance \cite{sophiya2025comprehensiveanalysispinnsvariants, 10.1093/imamat/hxae011, chuang2022experiencereportphysicsinformedneural}.\\
In this study, we provide insights into the performance of quantum neural networks (QNNs) in state-vector simulations in solving problems described by PDEs by integrating them into PINNs. We use the properties of QNNs as universal function approximators \cite{PerezSalinas2020} to learn the solution of a given PDE by training a quantum physics-informed neural network (qPINN). Previous studies \cite{dutta2024aqpinnsattentionenhancedquantumphysicsinformed, Berger2025, Siegl2025, e26080649, Sedykh_2024} indicate that in some cases using the potential of QNNs \cite{Abbas_2021, abbas2021effectivedimensionmachinelearning, Schuld_2021, Caro2022} in state-vector simulated qPINNs can lead to an improved performance, finding better trainability \cite{Siegl2025} and reduction of final loss \cite{Berger2025, e26080649, Sedykh_2024}. Advantages of this approach compared to other promising quantum approaches where the solutions are stored in a large state vectors is that no linearization of the PDE is necessary \cite{11344607, Liu_2021, SUCCI2024106148}, further information such as observations or objectives can be included, and that the readout problem can be avoided once the network is trained \cite{11344607, Liu_2021, SUCCI2024106148, PhysRevResearch.6.033257, PhysRevA.101.010301}. \\
In this paper, we show that the advantage of simulated qPINNs stems from their convergence rate enabling them to find a good approximation of the solution much faster - in terms of training epochs - than cPINNs, simultaneously explaining the apparent reduced final losses of qPINNs. Although both network types allow for comparable approximations after sufficient training epochs, the faster convergence of qPINNs has the potential to address the bottleneck of cPINNs. We show this by performing extensive training with up to $2\cdot10^4$ epochs for qPINNs and $1\cdot10^6$ epochs for cPINNs, taking the potentially highly complex loss landscape of PINNs into account \cite{wang2020understandingmitigatinggradientpathologies}. Moreover, we evaluate the conditions under which PINN solutions benefit from including parametrized quantum circuits. Two different network types are considered. The first type is a dense purely classical network, the second type is a quantum-classical (hybrid) architecture. A comprehensive overview of the performance of the qPINNs is provided by solving parametrized PDEs with generic boundary conditions. Information about the update of training data and adaptive loss weights are given in the Supplemental Material , together with further results, and training insights.

%\section{Methods}
%\subsection{Physics-informed neural networks}

%\subsection{System definition}\label{chapter:problem_definition}
\section{Methodology}
\subsection{Physics-informed neural networks}
In general, a PDE can be defined by 
\begin{align}\label{eq:general_pde}
    \mathcal{D}(u(t,x)) = 0
\end{align}
where $\mathcal{D}$ is a differential operator and $u(t,x)$ is a function depending on time $t$ and space $x$. Since the PDE itself only describes the change of quantities $u$ in space and time, some fixed information have to be given as well to guarantee a unique solution. These information are typically in the form of temporal and spatial boundary conditions $u_{t}$ and $u_{x}$
\begin{align}
    &u(t, x)&&= u_{t}(x), &&(t, x) \in [0] \times \Omega, \label{eq:general_boundaries_1}\\
    &u(t, x)&&= u_{x}(t,x), &&(t, x) \in [0, T] \times  \partial\Omega, \label{eq:general_boundaries_2}
\end{align}
where $\Omega$ and $\partial\Omega$ denote the spatial domain and its boundaries in $x$ (which can be multidimensional) and T the temporal domain. However, for PINNs, providing arbitrary fixed points, e.g. observations or objectives, works as well. To solve a PDE by using a neural network, the network takes temporal and spatial coordinates $t, x$ as inputs and outputs an approximation of the solution. Here, the network $f(t, x; \boldsymbol{\theta})$ can be any differentiable and parametrized network/function. If $\boldsymbol{\theta}$ were adjusted so that $f(t, x; \boldsymbol{\theta})=u$ is the optimal solution, Eqs. (\ref{eq:general_pde}), (\ref{eq:general_boundaries_1}) and (\ref{eq:general_boundaries_2}) would be given exactly. Therefore, a loss function $\mathcal{L}$ to be minimized can be defined as
\begin{align}\label{eq:pinn_loss_appendix}
    \mathcal{L}_{\boldsymbol{\theta}} &=  \mathcal{L}_{t} + \mathcal{L}_{x} + \mathcal{L}_{pde} \\
    \mathcal{L}_{t} &= \sum_{\mathclap{(t,x) \in [0]\times\Omega}} |f(t, x|\boldsymbol{\theta}) - u_{t}(x)|^2 \nonumber \\
    \mathcal{L}_{x} &= \sum_{\mathclap{(t,x) \in [0, T] \times \partial \Omega}} |f(t, x|\boldsymbol{\theta}) - u_{x}(t, x)|^2. \nonumber \\
    \mathcal{L}_{pde}&=  \sum_{\mathclap{(t,x) \in [0, T] \times \Omega}} |\mathcal{D}(f(t,x|\boldsymbol{\theta}))|^2. \nonumber \\ 
\end{align}
While the terms $\mathcal{L}_{t}$ and $\mathcal{L}_{x}$ are standard regression tasks, $\mathcal{L}_{pde}$ relies on the differentiability of $f(t,x|\boldsymbol{\theta})$, i.e., $\frac{\partial f(t,x|\boldsymbol{\theta})}{\partial t}$ which approximates $\frac{\partial u}{\partial t}$ for a parameter set $\boldsymbol{\theta}$, containing all the trainable parameters. This enables estimating the solution in the domains where no actual data, such as boundary conditions, is given. Since both, classical neural networks and quantum circuits,  are differentiable in that sense \cite{PhysRevA.99.032331}, they can be used to approximate the solution $u$ to a given PDE.
In this study, the temporal and spatial boundary conditions are treated equivalently (as they are mathematically similar) by defining $\mathcal{L}_{bounds}=\mathcal{L}_t + \mathcal{L}_x$. As mentioned before, $\mathcal{L}_{bounds}$ could even be replaced or supplemented with observations or objectives within the domain when using the PINNs to find a solution. For the training of these networks it turns out to beneficial to weight the two parts of Eq.~(\ref{eq:pinn_loss_appendix}) differently, effectively changing the loss landscape.
\begin{align}
    \mathcal{L}_{\boldsymbol{\theta}} &=  w_{bounds}\mathcal{L}_{bounds}+w_{pde}\mathcal{L}_{pde}. \label{eq:pinn_loss}
\end{align}
Based on this, $f(t,x|\boldsymbol{\theta})$ approximates $u$ by finding a set of parameters 
\begin{align}
  \boldsymbol{\theta}= \underset{\boldsymbol{\theta}^*}{\arg\min}  \; \mathcal{L}_{\boldsymbol{\theta}^*}. 
\end{align}

\subsection{System definition}
 Since temporal $u_t$ and spatial boundary conditions $u_x$ are incorporated into the loss term $\mathcal{L}_{bounds}$ by the same mathematical description (see Eq.~(\ref{eq:pinn_loss_appendix})), the temporal boundary conditions are considered representative. Without loss of generality the spatial boundary conditions are chosen depending on the temporal boundary conditions by fixing $u_x(t,x=0) = u_t(x=0)$ and $u_x(t,x=1) = u_t(x=1)$ over the whole temporal domain. We consider two distinct types of temporal boundary conditions and forcings. {\color{black}While these boundary conditions do not establish general applicability, they give an impression on the performance of qPINNs compared to cPINNs.} The set $'xsin'$ is used in the main part of this study:
\begin{align}
    &u^{xsin}_t(x) = \sin(3\pi x)\cdot x, \label{eq:xsin_1} \\
    &F^{xsin}(t)=\sin(4\pi t). \label{eq:xsin_2}
\end{align}
The results for a second set $'poly'$, based on polynomials, are presented mainly in the Supplemental Material :
\begin{align}
   &u^{poly}_t(x) = 50x(0.3-x)(0.6-x)(1.0-x), \label{eq:poly_1} \\
   &F^{poly}(t)=15t(0.4-t)(1.0-t). \label{eq:poly_2}
\end{align}
Regarding the PDE loss term $\mathcal{L}_{pde}$, we consider a broad variety of applications by introducing a parametrized PDE:
\begin{align}
    \mathcal{D}(L, N, F) = \frac{\partial u}{\partial t} - L \cdot \frac{\partial^2 u}{\partial x^2} + N \cdot u \frac{\partial u}{\partial x} - F(t)
    \label{eq:paraPDE}
\end{align}
with parameters $L$ and $N$ and forcing $F$. The operator $\mathcal{D}$ resembles Burgers' equation for $N=1$, for $N=0$ $\mathcal{D}$ becomes the heat equation. The parameter $L$ scales the linear term. The following values are taken into account:
\begin{align*}
L \in [0.01, 0.03, 0.1, 0.3, 1.0],\quad N  \in  [0.0, 1.0]. 
\end{align*}
We trained cPINNs and qPINNs for all combinations of boundary conditions and parameterized PDEs in order to judge their potential for solving real world problems.

\begin{figure}[t]
    %\centering
    %\resizebox{0.4\textwidth}{!}{%
    %\input{images/xsin_individual_example.pgf}
    %}
        \resizebox{0.45\textwidth}{!}{%
       \includegraphics[width=\linewidth]{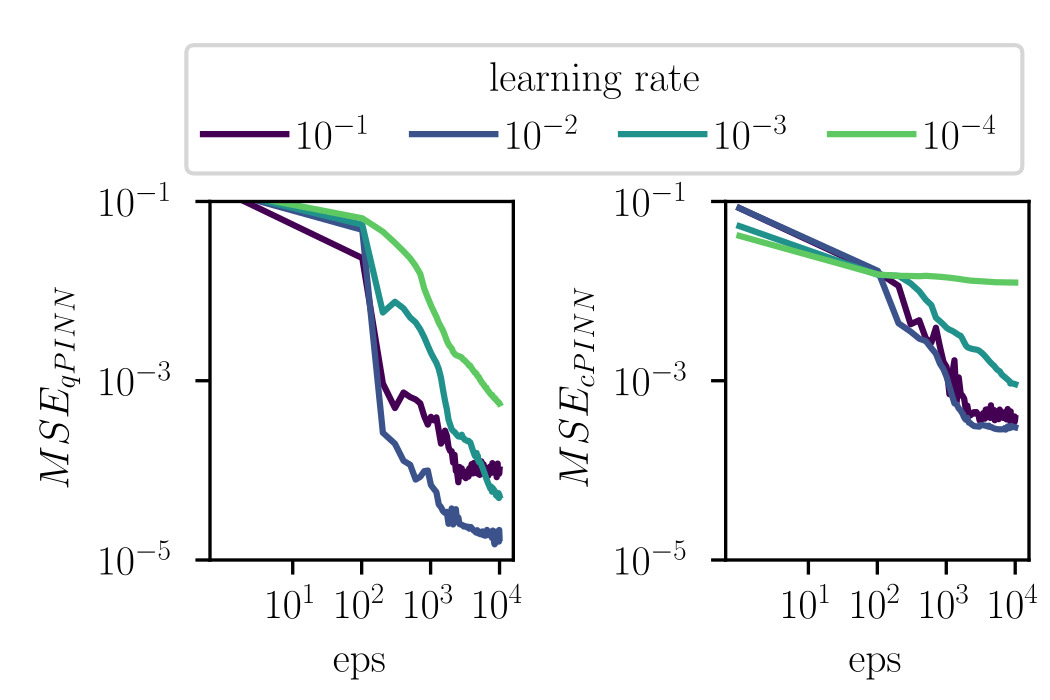}
       }
    \caption{Visualization of the training progress (mean squared error MSE as function of training epochs eps) for qPINNs (left) and cPINNs (right) for various learning rates. All trainings were conducted with 150 trainable parameters, a PDE defined by L = 0.1, N = 1.0 with $u^{xsin}_t$ and $F^{xsin}$ for boundary condition and forcing.}
    \label{fig:learning_rates}
\end{figure}
\subsection{Training settings}If not otherwise stated, $2\cdot 10^{4}$ epochs are used for training qPINNs and $1\cdot 10^{6}$ for cPINNs. {\color{black} Various learning rates are considered in Fig.~\ref{fig:learning_rates} in order to train both, qPINN and cPINN, with optimal training rates. The results indicate that a learning rate of $10^{-2}$, which is gradually decreased by a factor up to $0.1$ during training, is best for both architectures.} To evaluate Eq.~(\ref{eq:pinn_loss_appendix}), so-called collocation points are randomly sampled from $[0, T] \times \Omega$, $[0] \times\Omega$ and $[0, T] \times \partial \Omega$ to calculate the individual loss terms. In different training runs, 256, 512 or 1024 points are sampled for both, boundaries and domain by Sobol sampling \cite{SOBOL196786}. In each case four batches are used during training. Based on the calculated loss, the trainable parameters in the networks are updated by an Adam optimizer.

\subsection{Methods}
Two adaptations are made for the training of all PINNs in comparison to standard regression tasks. First, since training data can easily be sampled, the training data set is resampled whenever 
\begin{align}
\mathcal{L}^{validation}>1.1\cdot\mathcal{L}^{train}, \label{eq:update_data}
\end{align}
where $\mathcal{L}^{train}$ is the loss for seen, and $\mathcal{L}^{validation}$ for unseen data. This prevents the PINNs from overfitting and therefore enables investigating the full potential of each network (see Fig.~\ref{fig:methods} (a) in the Supplemental Material ).
Second, there is no unambiguous choice for the training weights in Eq. (\ref{eq:pinn_loss}) as they depend on the network and task. As it turns out, most ratios of weights eventually result in the same performance(see Fig.~\ref{fig:methods} (b) in the Supplemental Material ). However, a suitable choice of weights increases the convergence rate and therefore decreases the number of necessary training epochs. Although optimal weights could be tuned based on trained networks and known exact solutions, this not practicable in a real application. To avoid tuning the ratio before every training, an adaptive weighting strategy \cite{adaptiveWeights} is used. This leads to performances similar to those obtained with optimized weighting and, more importantly, the performances are independent of the choice of weights, minimizing the dependence on boundary conditions and PDE. 
\begin{figure*}[t]

    \begin{tikzpicture}

  \begin{scope}%[on background layer]
    \definecolor{color_q}{rgb}{0.702,0.357,0.941}   % 179/255, 91/255, 240/255
    \definecolor{color_e}{rgb}{0.357,0.941,0.604}   % 91/255, 240/255, 154/255
    \definecolor{color_v}{rgb}{0.941,0.729,0.357}   % 240/255, 186/255, 91/255
    
    \fill[gray!30,rounded corners=10pt] (-0.5,-0.0) rectangle (3.1,-3.2);
    \fill[gray!30,rounded corners=10pt] (13.45,-0.0) rectangle (17.0,-3.2);
    \fill[color_q!30,rounded corners=10pt] (3.1,-0.0) rectangle (13.45,-3.2);
    \fill[white] (3.7,-0.35) rectangle (8.3,-2.85);
    \fill[white] (9.1,-0.35) rectangle (11.9,-2.85);
    \fill[color_e!30] (3.7,-0.35) rectangle (8.3,-2.85);
    \fill[color_v!30] (9.1,-0.35) rectangle (11.9,-2.85);
  \end{scope}

    \draw [decorate,
    decoration={brace,amplitude=8pt,mirror}] (-0.5,-3.2) --  (3.1,-3.2) node[midway,below=10pt,black, align=center]{\textbf{classical input network} \\ $depth$ hidden layers \\ $width$ nodes per hidden layer};

    \draw [decorate,
    decoration={brace,amplitude=8pt,mirror}] (3.1,-3.2) --  (13.45,-3.2) node[midway,below=10pt,black, align=center]{\textbf{variational quantum circuit}\\$n_q$ qubits \\$2 \cdot n_{encoding}$ encoding layers};

    \draw [decorate,
    decoration={brace,amplitude=8pt,mirror}] (13.45,-3.2) --  (17.0,-3.2) node[midway,below=10pt,black, align=center]{\textbf{classical output network} \\ $depth$ hidden layers \\
    $width$ nodes per hidden layer};

    \draw [decorate,
    decoration={brace,amplitude=8pt}] (3.7,-0.2) --  (8.3,-0.2) node[midway,above=10pt,black, align=center]{encoding layer};

    \draw [decorate,
    decoration={brace,amplitude=8pt}] (9.1,-0.2) --  (11.9,-0.2) node[midway,above=10pt,black, align=center]{variational layer};

  % Neural network input nodes
  \foreach \m/\l [count=\y] in {t,x}
    \node [circle,draw, minimum size=16pt, inner sep=0pt] (input-\m) at (0,-0.8*\y-0.4) {$\m$};

  % First hidden layer nodes
  \foreach \m/\l [count=\y from 1] in {1/$\boldsymbol{\theta}$,2/$\boldsymbol{\theta}$,3/$\boldsymbol{\theta}$}
    \node [circle,draw,align=center, minimum size=16pt, inner sep=0pt] (hidden1-\m) at (1,-\y*0.8) {\l};

  % First hidden layer nodes
  \foreach \m/\l [count=\y from 1] in {1/$\boldsymbol{\theta}$,2/$\boldsymbol{\theta}$,3/$\boldsymbol{\theta}$}
    \node [circle,draw,align=center, minimum size=16pt, inner sep=0pt] (hidden2-\m) at (2,-\y*0.8) {\l};

  % Input to first hidden layer connections
  \foreach \i in {x,t}
    \foreach \j in {1,2,3}
      \draw [-] (input-\i) -- (hidden1-\j);

  \foreach \i in {1,2,3}
    \foreach \j in {1,2,3}
      \draw [-] (hidden1-\i) -- (hidden2-\j);

  \foreach \j/\k in {1/$i_{0}$, 2/$i_{j}$, 3/$i_{n_{q}}$}
    \draw [->] (hidden2-\j) -- node[above] {\k} (3,-\j*0.8);

    \draw[fill=gray!30, draw=black] (0.65,-2.7) rectangle (1.35,-0.5);
    \node at (1.0, -1.6) {$\cdots$};

  \node[inner sep=0, anchor=center, scale=1.0] (qcircuit) at (8.5, -1.6) {
    \fontsize{8pt}{8pt}\selectfont
    \Qcircuit @C=0.5em @R=1.0em {
      \lstick{\ket{0}} & \gate{R_{Y/X}(\makebox[1.5em][c]{$i_0$})} & \gate{R_{XYZ}(\boldsymbol{\theta})} & \ctrl{1} & \qw      & \targ     & \qw & \push{\cdots \; \;}  & \gate{R_{XYZ}(\boldsymbol{\theta})} & \ctrl{1} & \qw      & \targ     & \qw & \push{\cdots \; \;} &  \meter \\
      \lstick{\ket{0}} & \gate{R_{Y/X}(\makebox[1.5em][c]{$i_j$})}    & \gate{R_{XYZ}(\boldsymbol{\theta})} & \targ    & \ctrl{1} & \qw       & \qw & \push{\cdots \; \;}    & \gate{R_{XYZ}(\boldsymbol{\theta})} & \targ    & \ctrl{1} & \qw       & \qw & \push{\cdots \; \;}  & \meter \\
      \lstick{\ket{0}} & \gate{R_{Y/X}(\makebox[1.5em][c]{$i_{n_{q}}$})} & \gate{R_{XYZ}(\boldsymbol{\theta})} & \qw      & \targ    & \ctrl{-2} & \qw & \push{\cdots \; \;} & \gate{R_{XYZ}(\boldsymbol{\theta})} & \qw      & \targ    & \ctrl{-2} & \qw & \push{\cdots \; \;} & \meter \\
    }
  };

    % Post-circuit hidden layer nodes
  \foreach \m/\l [count=\y from 1] in {1/$\boldsymbol{\theta}$,2/$\boldsymbol{\theta}$,3/$\boldsymbol{\theta}$}
    \node [circle,draw,inner sep=2pt,draw,align=center, minimum size=16pt, inner sep=0pt] (hidden3-\m) at (14.5,-\y*0.8) {\l};

    % Post-circuit hidden layer nodes
  \foreach \m/\l [count=\y from 1] in {1/$\boldsymbol{\theta}$,2/$\boldsymbol{\theta}$,3/$\boldsymbol{\theta}$}
    \node [circle,draw,inner sep=2pt,draw,align=center, minimum size=16pt, inner sep=0pt] (hidden4-\m) at (15.5,-\y*0.8) {\l};

  \foreach \i in {1,2,3}
    \foreach \j in {1,2,3}
      \draw [-] (hidden3-\i) -- (hidden4-\j);

  % Output node
  \node [circle,draw, minimum size=16pt, inner sep=0pt] (output) at (16.5,-2*0.8) {$u$};

  % Draw arrows from circuit to third hidden layer nodes
  \foreach \j/\k in {1/$o_{0}$, 2/$o_{j}$, 3/$o_{n_{q}}$}
    \draw [->] (13.5, -\j*0.8) -- node[above] {\k} (hidden3-\j);

  % Draw arrows from third hidden layer to output
  \foreach \i in {1,2,3}
    \draw[->] (hidden4-\i) -- (output);

    \draw[fill=gray!30, draw=black] (15.15,-2.7) rectangle (15.85,-0.5);
    \node at (15.5, -1.6) {$\cdots$};

\end{tikzpicture}
    \caption{Schematic representation of a hybrid neural network architecture utilizing classical networks (gray) and  variational quantum circuits (purple). For encoding data into the circuit (left gray component) and decoding data from the circuit (right gray component) classical networks are used. These networks have inputs and outputs corresponding to the dimensions of the PDE and number of qubits $n_q$ in the quantum circuit. All parts of the hybrid network depend on trainable parameters $\boldsymbol{\boldsymbol{\theta}}$, which are optimized during training. The depth (hidden layers) and width (nodes per hidden layer) of the classical networks can be tuned for the specific task. Here, a three qubit circuit is visualized exemplary. The circuit consists of data encoding layers (green) and purely variational layers (orange). In encoding layers the outputs of the classical input network are encoded first in $R_Y$-gates, then in $R_X$-gates, each of them followed by a variational layer. Variational layers implement parameterized single-qubit rotations around all axis followed by entangling CNOT-gates. The number of encoding layers is defined by $n_{encoding}$, followed by adding variatonal layers, until the total number of trainable parameters for the network is reached. Expectation values of the Pauli-Z operator are used as outputs of the quantum circuit and input for the classical output network.}
    \label{fig:hybrid_network_visualization}
\end{figure*}
Incorporating the two algorithms in the training process enables reaching the full potential of the trained network as efficiently as possible.

\subsection{Networks} 
The purely classical network (cPINN) used in this study is a dense neural network. As activation function tanh is applied. The number of nodes is the same for all hidden layers. For a given number of trainable parameters $\boldsymbol{\theta}$ all possible numbers of hidden layers are tested to find the optimal configuration. The number of nodes per hidden layer is adjusted so that the desired number of trainable parameters is matched as closely as possible. After training the networks with each possible network depth, the depth resulting in the best approximation is chosen in the manuscript to represent the network size. {\color{black} More sophisticated classical architectures could also be explored, or the quantum circuits could be replaced with components that have strong non-linearities, such as classical reservoir computing systems \cite{Mammedov2022}. However, since this study focuses only on basic quantum circuits in a first step, we also focus on the corresponding basic classical networks and leave investigations of more complex quantum and classical networks for future work.}
\begin{figure}[t]
    \centering
    \setlength{\abovecaptionskip}{4pt}

    %  Row 1 
    \begin{minipage}[t]{0.46\textwidth}
        \centering
        \resizebox{\textwidth}{!}{\includegraphics[width=\linewidth]{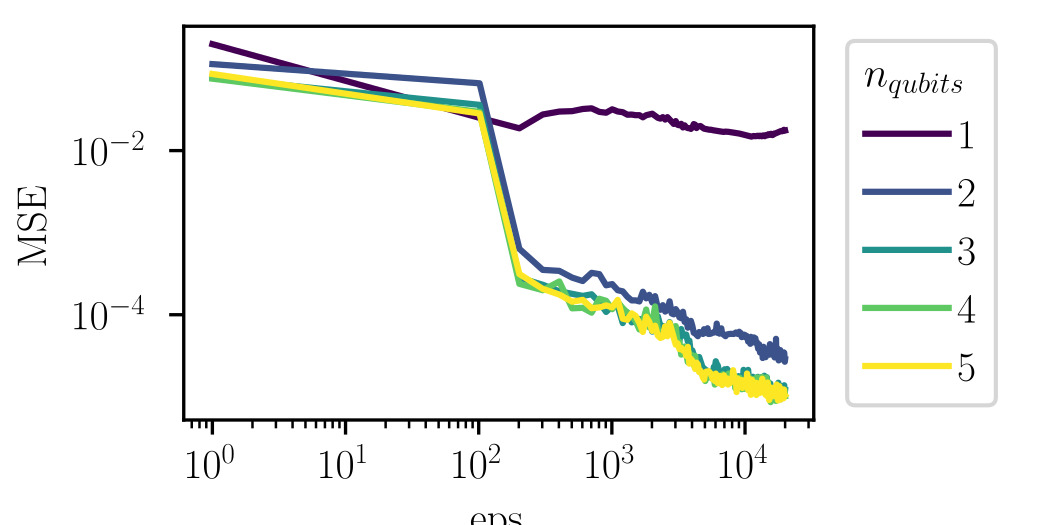}}
        \caption*{\textbf{(a)} Mean squared error (MSE) as a function of the number of training epochs for various numbers of qubits in the quantum circuit.}
    \end{minipage}
    \hfill
    \begin{minipage}[t]{0.46\textwidth}
        \centering
        \resizebox{\textwidth}{!}{\includegraphics[width=\linewidth]{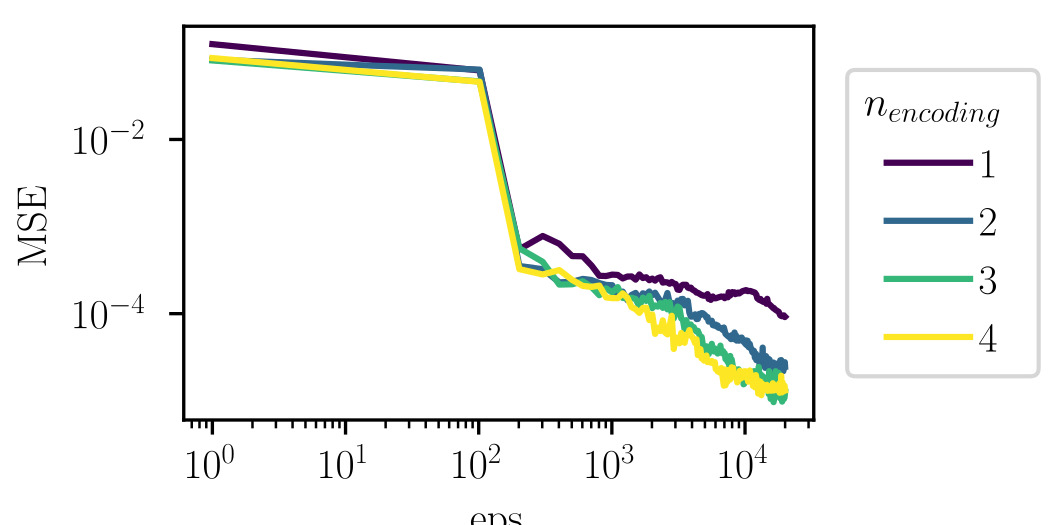}}
        \caption*{\textbf{(b)} Mean squared error (MSE) as a function of the number of training epochs for various numbers of encodings from the classical input layer into the quantum circuit.}
    \end{minipage}
    \hfill
    %  Row 2 
    \begin{minipage}[t]{0.46\textwidth}
        \centering
        \resizebox{\textwidth}{!}{\includegraphics[width=\linewidth]{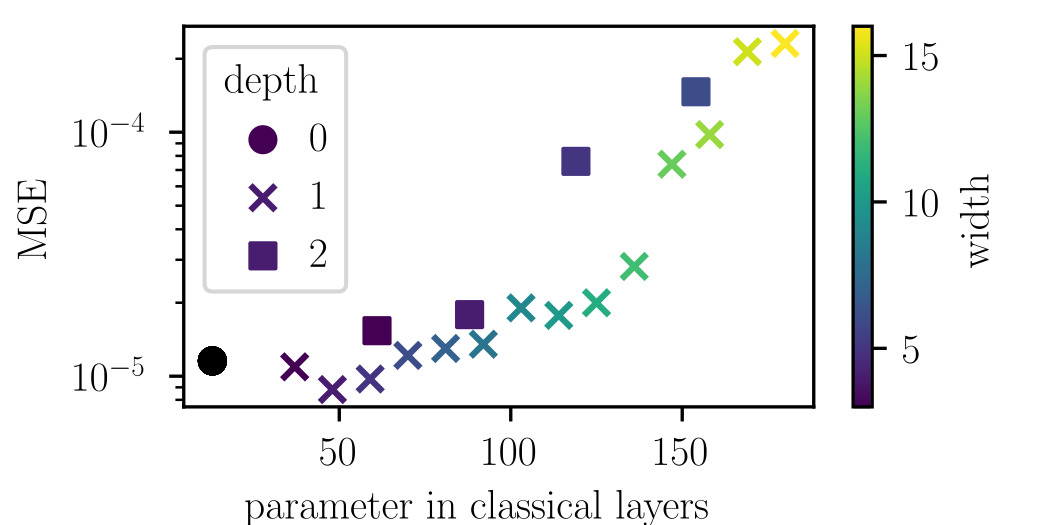}}
        \caption*{\textbf{(c)} Mean squared error (MSE) as a function of the number of parameter in the classical layers while keeping the total number of parameters fixed. More parameters in the classical layers means fewer parameters in the quantum circuit.}
    \end{minipage}

    \caption{Ablation study of a qPINN with default values $n_{qubits}=3$, $depth=1$, $width=6$, and data encoding after each encoding layer. The total number of trainable parameters is fixed at 200 and the number of data is 512 for each, boundary and PDE loss. The PDE is defined by $L=0.1, N=1.0$ with 'xsin' boundary conditions and forcing.}
    \label{fig:hybrid_network_study}

    \setlength{\abovecaptionskip}{10pt}
\vspace{-0.5cm}
\end{figure}

The qPINN consists of a QNN and two shallow neural networks for encoding and decoding data in and from the QNN, as illustrated in Fig.~\ref{fig:hybrid_network_visualization}. For implementing these networks, the python libraries pennylane \cite{bergholm2018pennylane}, equinox \cite{kidger2021equinox}, optax \cite{deepmind2020jax} and jax \cite{jax2018github} were used. Equinox enables easy integration of quantum circuits defined in pennylane in a differentiable network. Therefore the same training workflow, based on optax, can be applied to train all networks. jax accelerates the training process. For the simulation of the quantum circuit, state-vector simulation is used, which computes the full quantum state (all amplitudes) exactly—effectively corresponding to infinite shots—whereas shot-based simulation samples many probabilistic circuit executions to estimate measurement outcomes.

This architecture, used for incorporating quantum circuits into a hybrid network (qPINN) is inspired by two methods. First, the hardware efficient design of the quantum circuit itself is chosen as in \cite{Schuld_2021}. Second, to make efficient use of the Fourier frequencies, the input to the quantum circuit is transformed into a meaningful basis by a classical network \cite{PhysRevA.109.042421}. Consequently, it is not necessary to consider different encoding types \cite{PhysRevA.103.052416}. Finally, a second classical network maps the output of the quantum circuit to the solutions (Fig.~\ref{fig:hybrid_network_visualization}).\\
Fig.~\ref{fig:hybrid_network_study} shows the dependency on architecture parameters determined by a hyperparameter optimization. We found that the number of qubits does not matter as long as the number of trainable parameters remains the same (and as long as features can be encoded sufficiently, Fig.~\ref{fig:hybrid_network_study}(a)), which is in agreement with the spectrum size described in \cite{Schuld_2021}. Therefore, in the following, 3 qubits are used to reduce the computational effort to simulate the quantum circuit. The final mean squared error (MSE) benefits from encoding data as often as possible for a given number of parameters, as shown in Fig.~\ref{fig:hybrid_network_study} (b), again in agreement with \cite{Schuld_2021}. Here, each encoding means encoding in $R_Y$ and $R_X$, as illustrated in Fig.~\ref{fig:hybrid_network_visualization}. When optimizing the division of trainable parameters into classical and quantum parts (Fig.~\ref{fig:hybrid_network_study} (c)) it becomes obvious that the approximation benefits from using as many quantum resources as possible, consequently minimizing the parameters in classical layers. 

In the following, qPINNs with 0 and 1 classical layers are tested for each PDE. The better result is considered representatively for the performance of qPINNs for the specific PDE. This guarantees that a suitable encoding basis for each PDE is considered. The classical layers used for encoding and decoding have a width of 6 neurons. The quantum circuit consists of 3 qubits, and data is encoded as often the possible limited by the total number of trainable parameters. Finally, the circuit depth is adapted to the number of trainable parameters. 

\begin{figure}[t]

    \resizebox{0.45\textwidth}{!}{%
        \includegraphics[width=\linewidth]{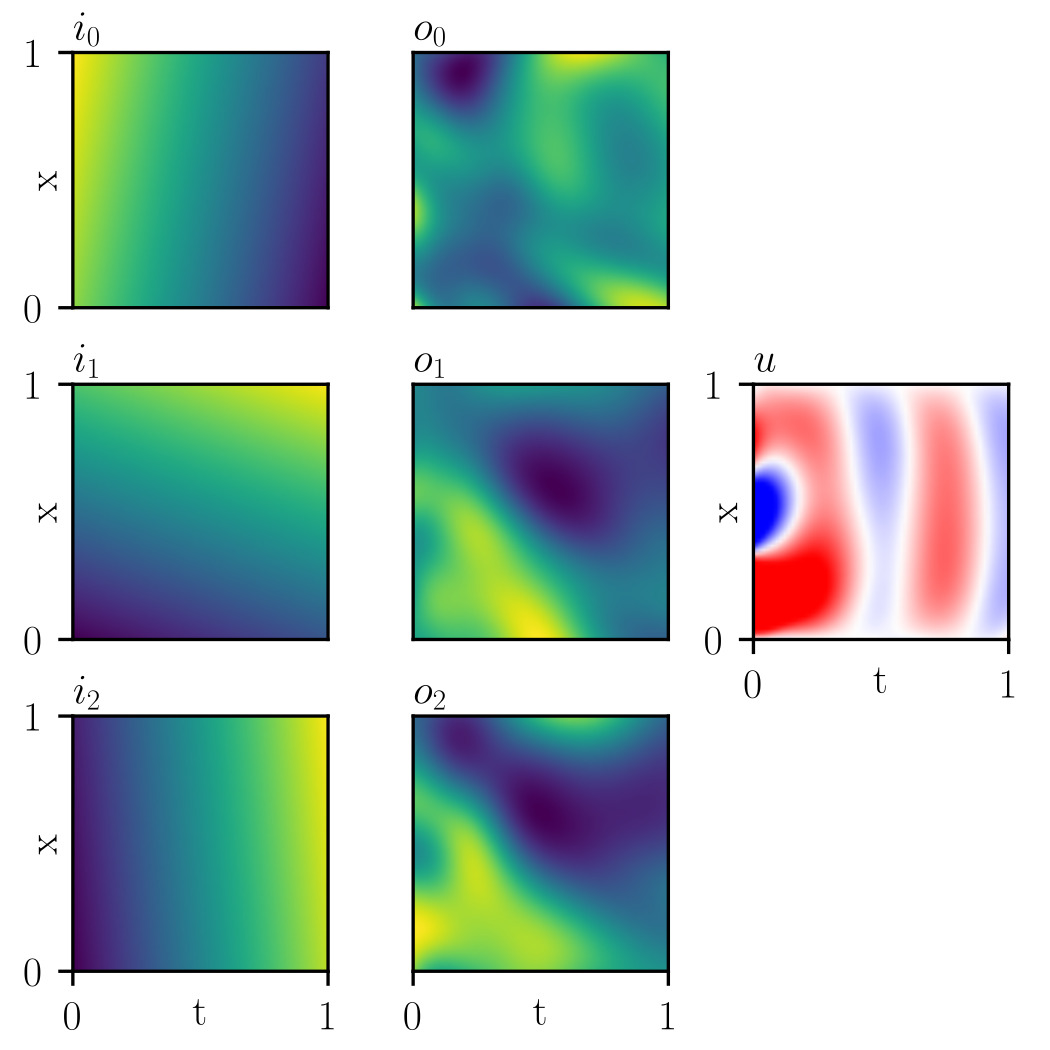}
    }
    \caption{Individual outputs of the hybrid architecture (as visualized in Fig.~\ref{fig:hybrid_network_visualization})  for a hybrid network based on a three-qubit quantum circuit. The total number of trainable parameters in the circuit is 200 and the number of training points is 512 for each, boundary and PDE loss. The considered PDE uses $L=0.1, N=1.0$ with 'xsin' boundary conditions and forcing. $i_0$ and $i_1$ represent the output of the classical input network. Since this is a qualitative visualization, no actual values corresponding to the colors are provided.}
    \label{fig:individual_output_hybrid}
\end{figure}

To illustrate the contribution of each of the three network parts from Fig.~\ref{fig:hybrid_network_visualization}, classical input network, quantum circuit, classical output network, their individual outputs are presented in Fig.~\ref{fig:individual_output_hybrid}. In this example, $i_0$ and $i_2$ tend to encode the $t$-coordinate, while $i_1$ encodes the sum of the $t$-coordinate and $x$-coordinate. This relates to the solution, which shows more variance across time than space, especially for later times. Obviously this can change significantly when the domain becomes much more complex than the square domain considered here, consequentially requiring more qubits and larger classical networks to find and encode a suitable basis. In the present case, the quantum circuit outputs $o_0$, $o_1$, and $o_2$ are based on the learned basis, which seems to be an efficient decomposition of the solution $u$ to be combined by the second classical network.

\subsection{Training \& evaluation}
The training process of both, cPINNs and state-vector simulated qPINNs, with the same number of trainable parameters, is exemplarily visualized in Fig.~\ref{fig:training_example}. Due to the adaptive weight balancing, the loss is no suitable performance measure for the approximation anymore. For this reason, the mean squared error (MSE, calculated every $100$ epochs) between the exact solution obtained by numerical simulation and the PINN approximation is used representatively. The MSE is calculated on a validation dataset of the same size as the training dataset. Although both networks reach a similar total MSE, the qPINN needs roughly 10 times fewer training epochs. For some approximation accuracies ($\mathrm{MSE} \sim 10^{-4}$), the qPINN is close to 100 times faster in terms of training epochs. Consequently, when training for a fixed number of epochs, the qPINN approximation is more accurate, in the presented instance around 50 times. However, eventually a similar MSE is reached, which is denoted as the ``general accuracy limit''. To compare qPINNs and cPINNs, the ratio of epochs $r_{eps}$ to reach a certain MSE and the ratio of the reached MSEs $r_{mse}$ after training for a fixed number of epochs is calculated, as illustrated in Fig.~\ref{fig:training_example}:

\begin{figure}[t]
    %\centering\href{}{}
    %\resizebox{0.4\textwidth}{!}{%
    %\input{images/xsin_individual_example.pgf}
    %}
        \resizebox{0.45\textwidth}{!}{%
       \includegraphics[width=\linewidth]{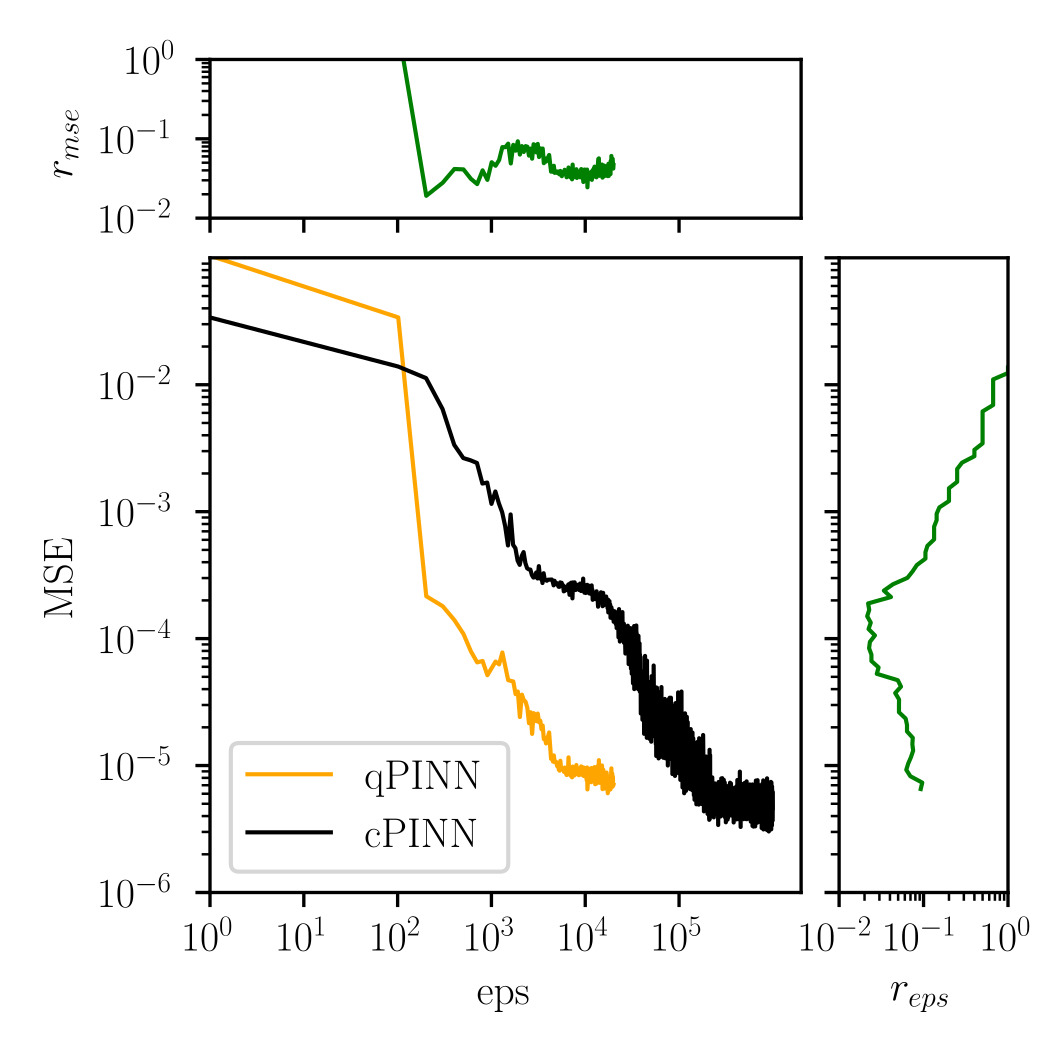}
       }
    \caption{Visualization of the median MSE over 10 training runs as a function of the number of training epochs for networks with 250 parameters trained on 1024 training points, for a PDE defined by $L=0.1, N=1.0$ with $u_t^{xsin}$ and $F^{xsin}$ for boundary condition and forcing. The epoch ratio to reach a certain MSE (right) and MSE ratio per epoch (top) are also shown. MSEs are calculated every 100 epochs.}
    \label{fig:training_example}
\end{figure}

\begin{align}
    r_{eps}=\frac{eps_{qPINN}}{eps_{cPINN}}, \quad r_{mse} = \frac{MSE_{qPINN}}{MSE_{cPINN}},
\end{align}
therefore, smaller $r_{eps}$ and $r_{mse}$ indicate advantages of the noise-free qPINN. 

To investigate the cause of the improvement in training, we systematically explore all possible combinations of the boundary conditions and PDEs described above. The number of training points for both, boundary loss and PDE loss, is also varied for each combination taking values of 256, 512 or 1024. Due to the inherent variability arising from the random sampling of collocation points and trainable parameters, each PINN's performance is characterized by the median outcome across 10 independent training runs.
{\color{black} We used architectures of the same size in terms of trainable parameters, optimized individual network architectures for each problem instance while making sure, that learning rates are optimal for both network types. This aims to investigate basic properties of classical and quantum PINNs under fair conditions. Since this study focuses on exploring the fundamental potential of qPINNs, shot noise is not considered. However, for practical applications, its impact must be analyzed and accounted for. The required number of shots might reduce the observed advantage of qPINNs in comparison to cPINNs, even though the small number of required qubits should make efficient parallelization possible. }

%\section{Results}
\section{Results}%
\begin{figure}[t]
    %\centering
    %\resizebox{0.4\textwidth}{!}{%
    %\input{images/xsin_individual_example.pgf}
    %}
        \resizebox{0.45\textwidth}{!}{%
       \includegraphics[width=\linewidth]{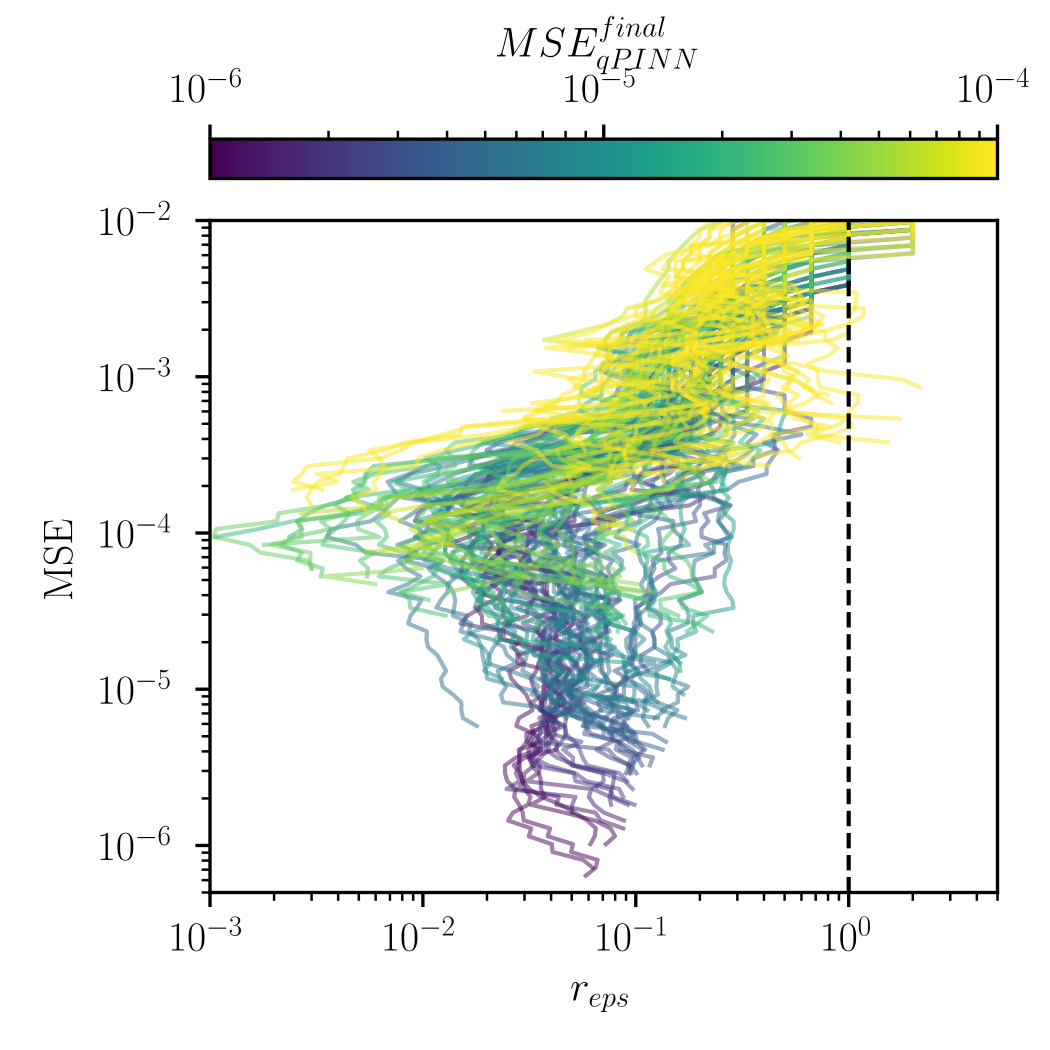}
       }
    \caption{Epochs needed to reach a given MSE when training qPINNs with respect to the epochs needed to reach the same MSE when training cPINNs. The figure presents an extension of the plot on the right in Fig.~\ref{fig:training_example}, covering all tested combinations of PDEs, numbers of collocation points, and trainable parameters. Line colors indicate the MSE of the qPINN after $2\cdot 10^4$ epochs, after which training of the qPINN was stopped. The dashed line marks where cPINNs and qPINNs perform equally.}
    \label{fig:convergence_rate_ratio}
\end{figure}
\begin{figure}[t]
    %\centering
    %\resizebox{0.4\textwidth}{!}{%
    %\input{images/xsin_individual_example.pgf}
    %}
        \resizebox{0.45\textwidth}{!}{%
       \includegraphics[width=\linewidth]{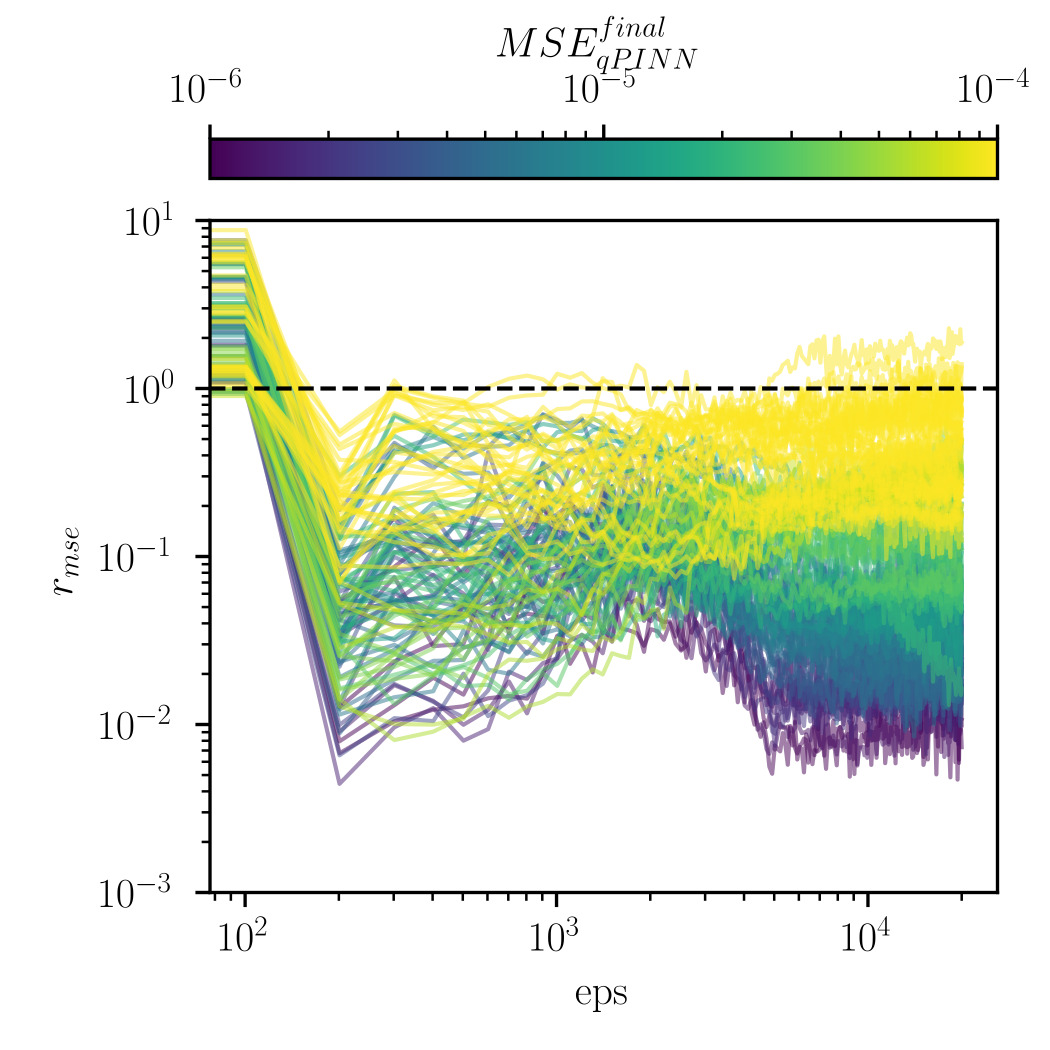}
       }
    \caption{MSE reached by qPINNs after training a given number of epochs (eps) with respect to the MSE reached by cPINNs after training for the same number of epochs. The figure presents a summary of the plot on the top in Fig.~\ref{fig:training_example}, covering all tested combinations of PDEs, numbers of collocation points, and trainable parameters. Line colors indicate the final MSE of the qPINN for the corresponding PDE. The dashed line marks where cPINNs and qPINNs perform equally.}
    \label{fig:mse_ratio_xsin}
\end{figure}

To understand the impact of incorporating quantum circuits into PINNs when solving PDEs, the training results for all possible combinations of PDEs (defined by $L$ and $N$), numbers of trainable parameters (100, 150, 200, 250) and number of training data (256, 512, 1024) are evaluated as exemplarily illustrated in Fig.~\ref{fig:training_example} and summarized in Fig.~\ref{fig:convergence_rate_ratio} and Fig.~\ref{fig:mse_ratio_xsin}. Results for further boundary conditions are provided in Fig.~\ref{fig:poly_results} in the Supplemental Material .

Fig.~\ref{fig:convergence_rate_ratio} corresponds to the right plot in Fig.~\ref{fig:training_example}, visualizing the ratio of epochs needed by qPINNs and cPINNs to reach certain MSEs. Some $r_{eps}$ go down to $10^{-3}$ during training, meaning that the qPINN needed a factor of 1000 fewer epochs to reach a given MSE. Most training runs reach an epoch ratio of $\sim10^{-2}$ before increasing again once both networks approach the general accuracy limit. As this happens earlier when the general accuracy limit is larger (bright curves) the epoch ratio is not reduced as much during training. A larger general accuracy limit is due to PDEs that are hard to solve in combination with insufficient small numbers of trainable parameters in the network. In general, the specific PDE does not appear to have a direct influence, but rather affects the general accuracy limit based on the number of parameters, which in turn is the main influence on the epoch ratio during training. This suggests that qPINNs appear to be equally beneficial for all PDEs considered as long as the number of trainable parameters is sufficient.
The training benefits from including quantum circuits in the networks not only from the perspective of epochs needed for achieving a specific MSE. There is also an improvement in terms of MSE when training for a fixed number of epochs (Fig.~\ref{fig:mse_ratio_xsin}). As with the epoch ratio, the MSE ratio strongly depends on the general accuracy limit. While for larger general accuracy limits qPINNs perform only slightly better (brighter lines in Fig.~\ref{fig:convergence_rate_ratio}), for small general accuracy limits the MSE of qPINNs is up to 100 times smaller than the MSE of cPINNs in the same training epoch (darker lines in Fig.~\ref{fig:convergence_rate_ratio}). Nevertheless, eventually all $r_{mse}$ seem to increase back to $1$ when approaching the general accuracy limit, if they were trained sufficiently long.

In summary, the general accuracy limit defines the potential of a qPINN compared to a cPINN of the same size. While a higher general accuracy limit, i.e., a large mean squared error remaining after training, restricts the possibility of improvement, a smaller general accuracy limit enables superior performance. This is not only true for the boundary conditions 'xsin' shown here, but applies also for the 'poly' boundary conditions presented in the Supplemental Material .

\begin{figure}[t]
    %\centering
    %\resizebox{0.4\textwidth}{!}{%
    %\input{images/xsin_individual_example.pgf}
    %}
        \resizebox{0.45\textwidth}{!}{%
       \includegraphics[width=\linewidth]{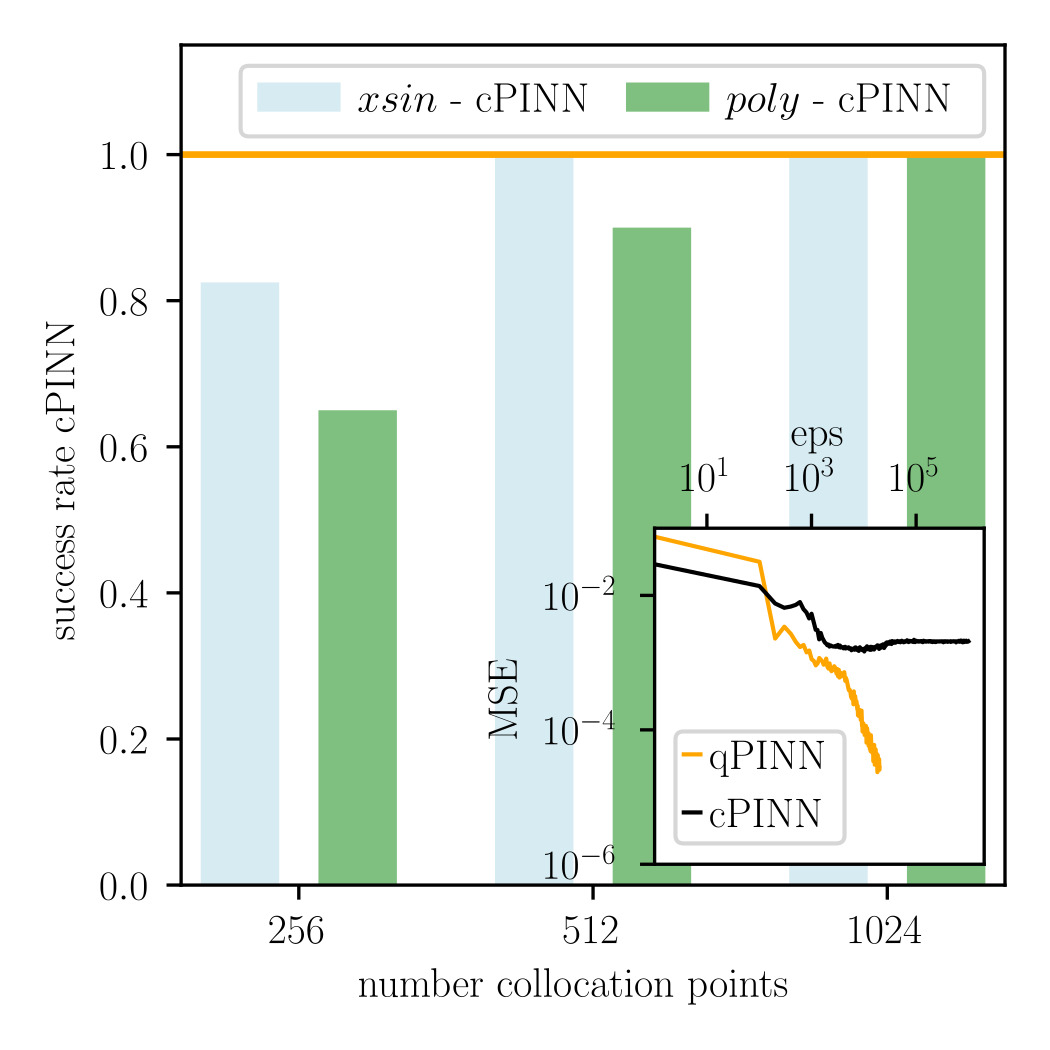}
       }
    \caption{Success ratio of training cPINNs with respect to qPINNs for various numbers of collocation points and types of boundary conditions and forcings (Eqs. \ref{eq:xsin_1}, \ref{eq:xsin_2} and Eqs. \ref{eq:poly_1}, \ref{eq:poly_2}, cPINNs in blue/green for xsin resp. poly boundary conditions). Trainings are considered not successful when the network gets stuck in a local minimum, as visualized in the inset. When the number of collocation points is sufficiently high, the training is always successful. Even though the training of qPINNs can be slowed down by a reduction of training data, they did not get stuck for any of the tests, as illustrated by the orange line.}
    \label{fig:success_ratio}
\end{figure}

\subsection{Training data size and solution stability}
Not only does the training of qPINNs converge in fewer training epochs than the training of cPINNs, their training is also more stable when the number of collocation points is small in relation to the problem complexity. In general, reducing the amount of training data can slow down the training, but the final performance does not seem to be impacted, as long as the network does not get stuck in a local minimum. As visualized in Fig.~\ref{fig:success_ratio}, for large training datasets all instances of cPINNs and qPINNs are trained successfully. However, when reducing the number of collocation points, the success ratio for cPINNs is reduced, which can be caused by the cPINNs getting stuck in a local minimum, limiting their performance. Meanwhile, qPINNs are trained successfully even with the smallest number of  collocation points tested.

\begin{figure}[t]
    \centering
    \includegraphics[width=0.49\textwidth, trim=0cm 8.5cm 0cm 2cm, clip]{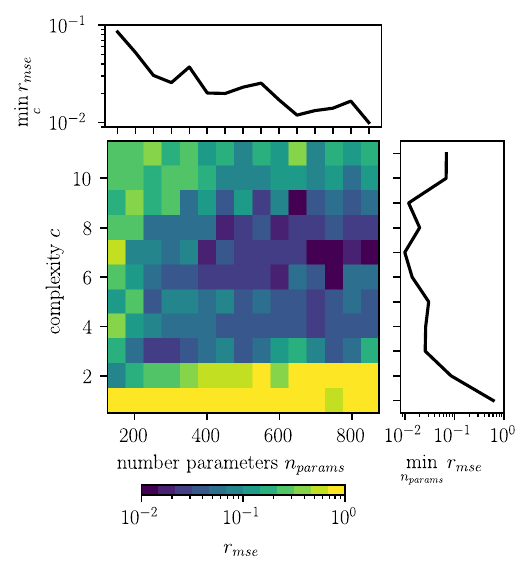}
    \caption{Ratio of MSEs reached by qPINNs and cPINNs after $5\cdot10^3$ training epochs for various complexities and numbers of trainable parameters. The figure above the main plot shows the smallest ratio for each number of trainable parameters across all complexities, while the figure to the right shows the smallest ratio for each complexity across all parameter counts.}
    \label{fig:scaling_1}
\end{figure}

\begin{figure}[t]
        \centering
        \includegraphics[width=0.49\textwidth, trim=0cm 1cm 0cm 12cm, clip]{images/improvement_4.pdf}
        \caption{The ratio of MSEs reached by qPINNs and cPINNs as function of the MSE reached by qPINNs. The data points correspond to the performance evaluated at various training epochs ranging from $1\cdot10^2$ to $5\cdot10^3$ and for networks with 150 to 850 trainable parameters. The system’s complexity is indicated by the color coding.}
        \label{fig:scaling_2}
\end{figure}

\subsection{Scaling with complexity}
To get an impression of how the benefits of qPINNs might scale with problem complexity, trainings with parameterized boundary conditions $u^c_t$ and forcings $F^c$ were conducted:
\begin{align}
    u^c_t(x) = \sin(c \cdot \pi x)\cdot x, \quad F^c(t)=\sin(c \cdot \pi t),
\end{align}
where the complexity increases with the parameter $c$. The PDE is defined by $L=0.1$, $N=1.0$, the qPINN is based on 4 qubits, data is encoded as often as possible. The classical layers consists of 1 hidden layer with 6 nodes to enable finding a sufficient basis for all complexities. The scaling behavior of the convergence rate differences of qPINNs compared to cPINNs in terms of the MSE ratio is visualized in Fig.~\ref{fig:scaling_1} and  Fig.~\ref{fig:scaling_2}. Fig.~\ref{fig:scaling_1} illustrates that the efficiency of qPINNs (measured by the MSE ratio) scales with the problem complexity. This is visible by the decreasing minimal MSE ratio with increasing complexity (Fig.~\ref{fig:scaling_1}, right), when a sufficient number of trainable parameters is provided. However, for complexities above 9, the number of trainable parameters is not sufficient yet, allowing only for rough solution approximations and therefore larger minimal MSE ratios. The same is valid for the number of parameters, the MSE ratio decreases when increasing the number of trainable parameters when the problem is sufficiently complex (Fig.~\ref{fig:scaling_1}, top). 
Also, Fig.~\ref{fig:scaling_2} indicates that the minimal MSE ratio shifts to smaller values with increasing complexity when providing the required number of parameters. This is evident from the decreasing MSE ratios as complexity increases. As mentioned earlier, for higher complexities ($c>9$, yellow data), the number of trainable parameters is still insufficient to achieve low MSEs and thus low $r_{mse}$. Additionally, for a fixed complexity, the relationship between the qPINN MSE (after various numbers of training epochs) and the respective MSE ratio appears to follow a power-law behavior.
Although no generalization is possible, the results indicate that the advantage of qPINNs due to the reduced number of training epochs might increase with the problem complexity. 

\section{Discussion}%  
The presented results demonstrate that state-vector simulated qPINNs can be trained in significantly fewer training epochs than cPINNs. {\color{black} The contribution of the variational quantum circuits as the primary driver of the observed behavior is already visible in Fig.~\ref{fig:hybrid_network_study} (c), which shows that, for a fixed number of trainable parameters, the performance improves with a larger proportion of quantum components.} The reduction of the training epochs depends on the ability of the network to find a well fitting solution eventually, characterized by the general accuracy limit. Potentially, this is caused by the number of approximate solutions that achieve a given MSE. Considering the space of approximate but imperfect solutions (for fixed boundary conditions and PDE), only a small subset corresponds to an error near the general accuracy limit. The lower the limit, the smaller this subset becomes. Consequently, finding parameters corresponding to this subset during training becomes more demanding. Hence, the apparently efficient exploration of the loss landscape by qPINNs is more beneficial the lower the general accuracy limit. The larger the general accuracy limit becomes, the larger the number of possible, imperfect solutions in the loss landscape. Since there are as many configurations of trainable parameters corresponding to these solutions, the training becomes less demanding and the efficient exploration by qPINNs less beneficial. The more efficient exploration of the loss landscape is also emphasized by the more stable training of qPINNs when little training data is available.
{\color{black} Future work will further investigate this explanation, identify which components of the quantum circuits drive the observed behavior, and examine the effects of boundary conditions and the governing PDEs.} Another important property is the scaling of the presented effects in real world applications with e.g., network capacity, more complex domains or higher dimensions. Especially when considering more difficult problems in higher dimensions an even more complex loss landscape can be expected and more training data needed, resulting in a high potential for qPINNs because of their efficient exploration of loss landscapes (as indicated in Fig.~\ref{fig:scaling_1} and Fig.~\ref{fig:scaling_2}) and trainability on few training data. Consequently, qPINNs could become useful when dealing with real world problems, involving complex geometries, coupled PDEs and multiscale problems. For a meaningful application of qPINNs in such problems, a much higher expressibility is needed, therefore requiring larger networks in terms of trainable parameters. As these networks can not be trained by state vector simulations on classical computers, a transition to real hardware is necessary to investigate the scaling behavior, especially in the possible presence of barren plateaus. For this, shot noise, errors in current quantum devices, and their impacts on the results have to be taken into account.

\section{Conclusion}% 
The paper shows that integrating variational quantum circuits in PINNs is beneficial for their training in state vector simulations. By using classical layers for encoding and decoding data into quantum circuits, a suitable basis can be found to access their full capabilities. As this leads to considerably improved training efficiency due to a larger convergence rate in terms of training epochs, qPINNs offer great potential for the application in data-driven, complex problems, not only outperforming training-intensive cPINNs but possibly also iterative numerical methods. Therefore, it is valuable to investigate whether e.g., climate models can be enhanced by integrating observational data and parametrizations directly into the loss function while solving the underlying PDEs by qPINNs. This approach is promising to address the computationally costly parameter tuning process, thereby replacing separate, explicit optimization procedures.

\begin{acknowledgments}
\section{Acknowledgments}This project was made possible by the DLR Quantum Computing Initiative and the Federal Ministry of Research, Technology and Space; \url{qci.dlr.de/projects/klim-qml}. V.E. was additionally supported by the
Deutsche Forschungsgemeinschaft (DFG, German Research Foundation) through the Gottfried Wilhelm Leibniz Prize awarded to Veronika Eyring (Reference No. EY 22/2-1). This work used resources of the Deutsches Klimarechenzentrum (DKRZ) granted by its Scientific Steering Committee (WLA) under project ID bd1179.
\end{acknowledgments}

%\nocite{*}

\bibliography{literature.bib}

\clearpage
\appendix

\section{Supplemental Material}
\setcounter{page}{1}
\begin{figure}[t]
    \centering
    \setlength{\abovecaptionskip}{3pt}

    %  Row 1 
    \begin{minipage}[t]{0.49\textwidth}
        \centering
                \resizebox{0.8\textwidth}{!}{%
                \hspace*{-1cm}
       \includegraphics[width=\linewidth]{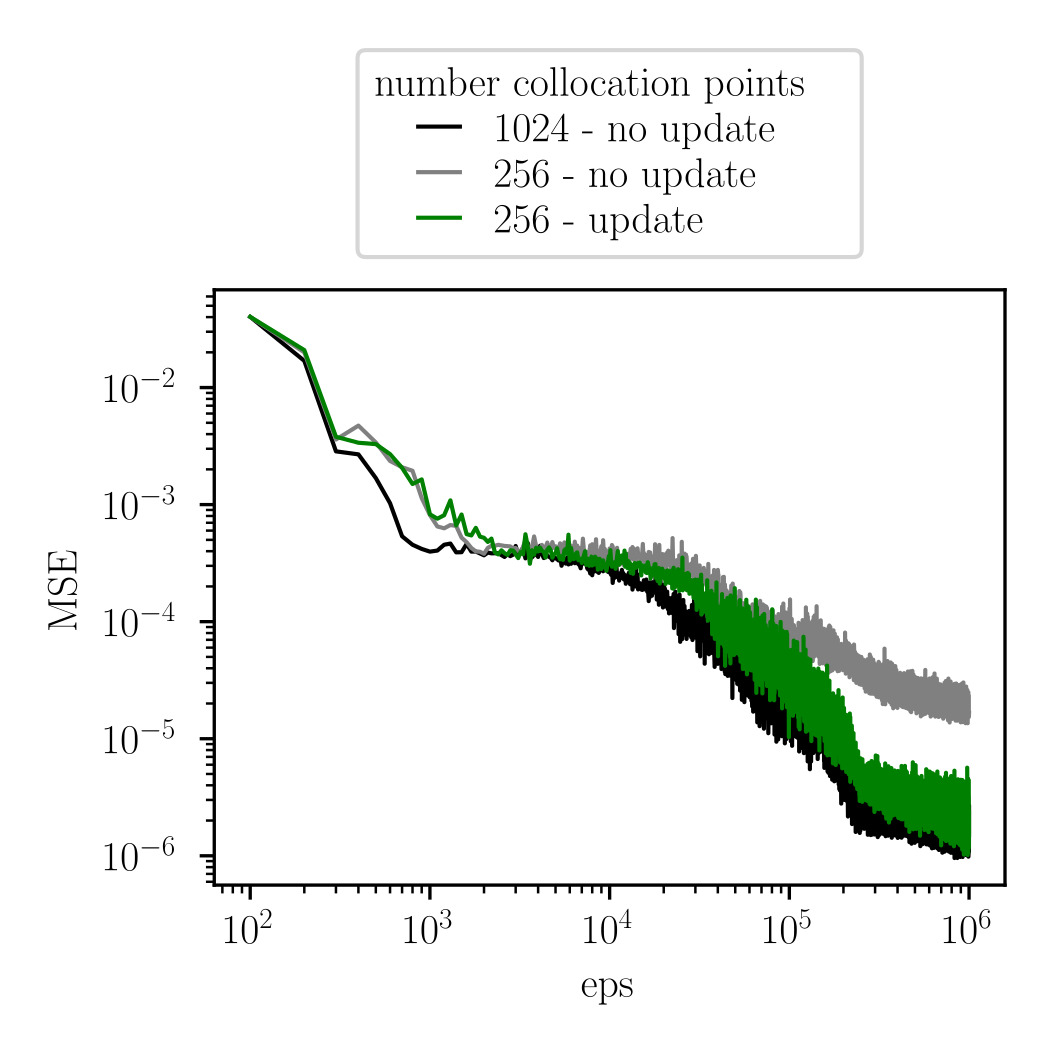}
       }
        \caption*{\textbf{(a)} Mean squared error (MSE) for cPINNs calculated with 1024 and 256 collocation points, with and without updating the training data}
    \end{minipage}
    \hfill
    \begin{minipage}[t]{0.49\textwidth}
        \centering
               \resizebox{0.8\textwidth}{!}
               {
                \hspace*{-1cm}
               \includegraphics[width=\linewidth]{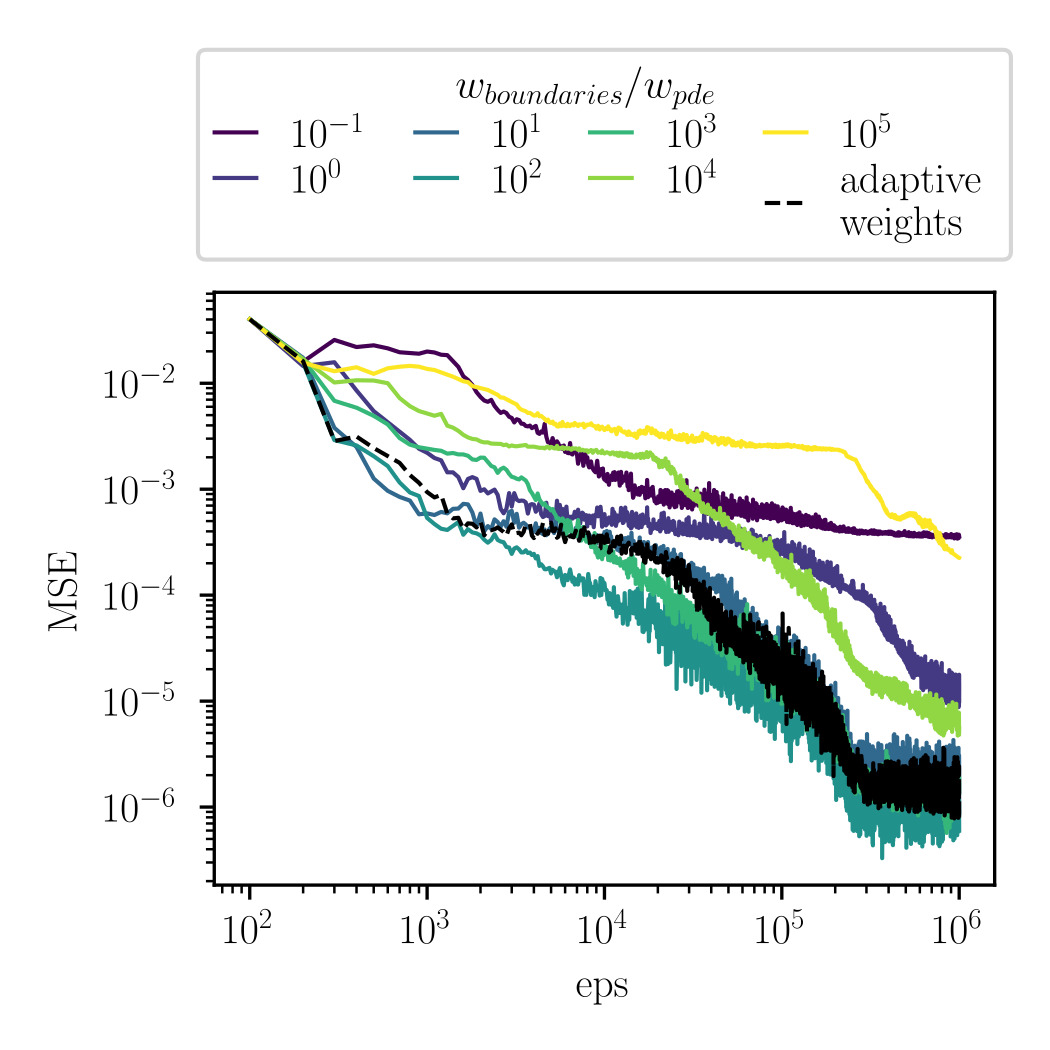}
        }
        \caption*{\textbf{(b)} Mean squared error (MSE) for cPINNs calculated for various numbers of weight ratios $w_{boundaries}/w_{pde}$, resp. with an adaptive weighting algorithm as function of training epochs eps.}
    \end{minipage}

    \caption{Effect of updating training data (a) and choice of weights (b). All trainings were conducted with 250 trainable parameters, a PDE defined by $L=0.1$, $N=0.0$ with $u_t^{xsin}$ and $F^{xsin}$ for boundary condition and forcing.}
    \label{fig:methods}

    \setlength{\abovecaptionskip}{10pt}
    \vspace{-0.2cm}
\end{figure}

\subsection{Updating training data}
To enable networks to find the optimal solution and thus reach their full potential, sufficient training data is required. While for regression tasks the training data is often limited, this can be circumvented in PINNs by generating additional collocation points. One strategy is to use a fixed larger data set throughout the training process. Alternatively, especially for high-dimensional or complex systems, smaller sets can be used and updated once the network has effectively learned the current data and begins to overfit, as described in Eq.~\ref{eq:update_data}. The options are compared in Fig.~\ref{fig:methods}~(a). While a larger number (1024) of collocation points enable the network to to reach a MSE of $10^{-6}$, the smaller number (256) of only reaches a MSE of $10^{-5}$. By updating the training data, the network manages to reach a MSE of $10^{-6}$ while reducing the costs of training. Furthermore, the is no need to specify the exact number of collocation points. 
Using 256 collocation points without update limits the solution accuracy compared to 1024 collocation points. When using 256 collocation points, the accuracy is the same as when using 1024, enabling the utilizing the networks full potential.

\subsection{Adaptive weights}
The effect of using adaptive weights instead of fixed weights in Eq.~\ref{eq:pinn_loss} and Eq.~\ref{eq:pinn_loss_appendix}, as mentioned in the main part, is visualized in Fig.~\ref{fig:methods} (b). The adaptive weights are defined as the inverse of the norm of the respective loss gradient with respect to the trainable parameters \cite{adaptiveWeights}. As illustrated by the dashed lines, this sufficiently resembles an optimal choice of weights by adaptively balancing the loss terms. While using adaptive weights may not always be the optimal configuration, it eliminates the need to tune fixed weights, a process that is impractical in real-world applications, since the true solution would have to be known. More important, it also prevents the results from being dependent on the PDE under consideration, as they might benefit from different fixed weights.

\vfill

%\clearpage

%\subsection{Architectures}

\clearpage
\subsection{Results for further boundary conditions and forcings}
\begin{figure}[t]
    \centering
        \resizebox{0.45\textwidth}{!}{\includegraphics[width=\linewidth]{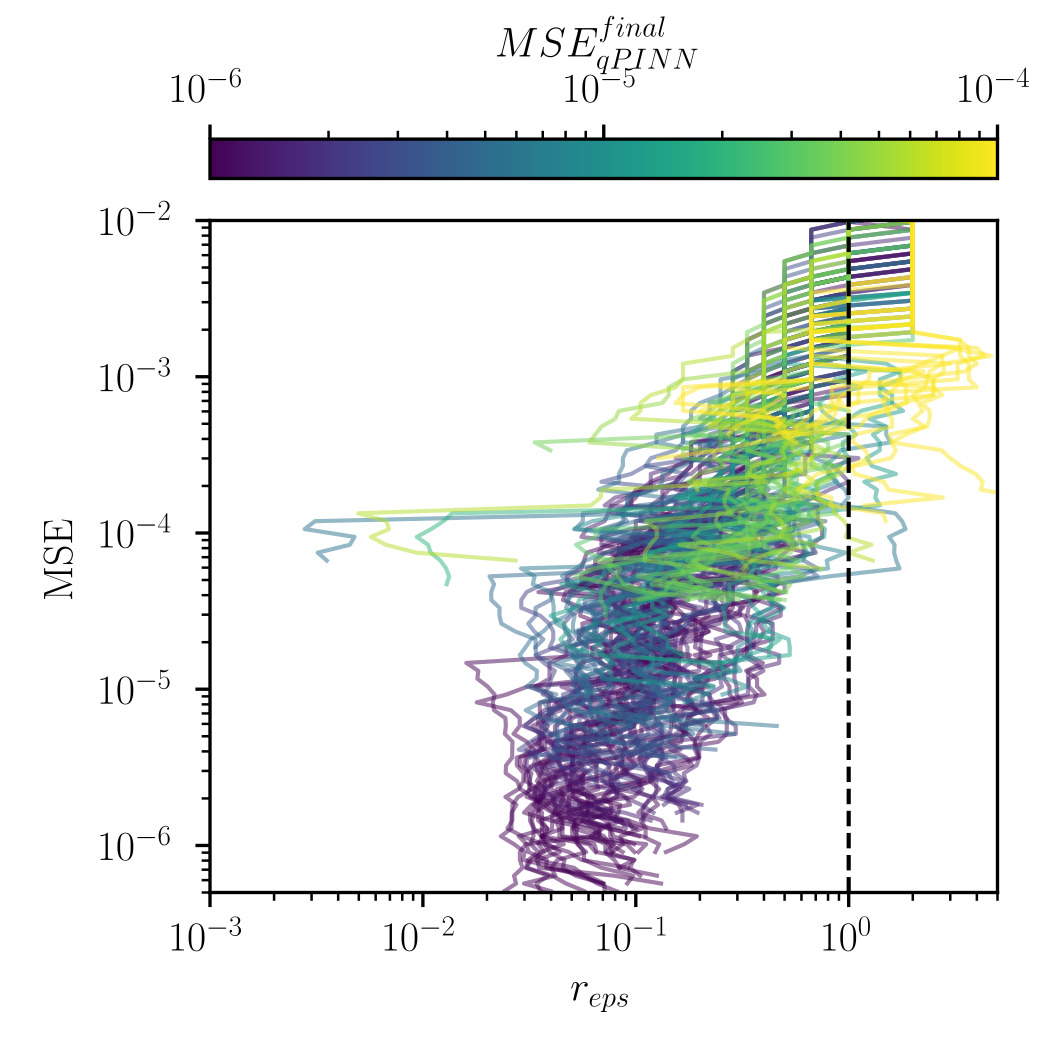}
        }

        \resizebox{0.45\textwidth}{!}{\includegraphics[width=\linewidth]{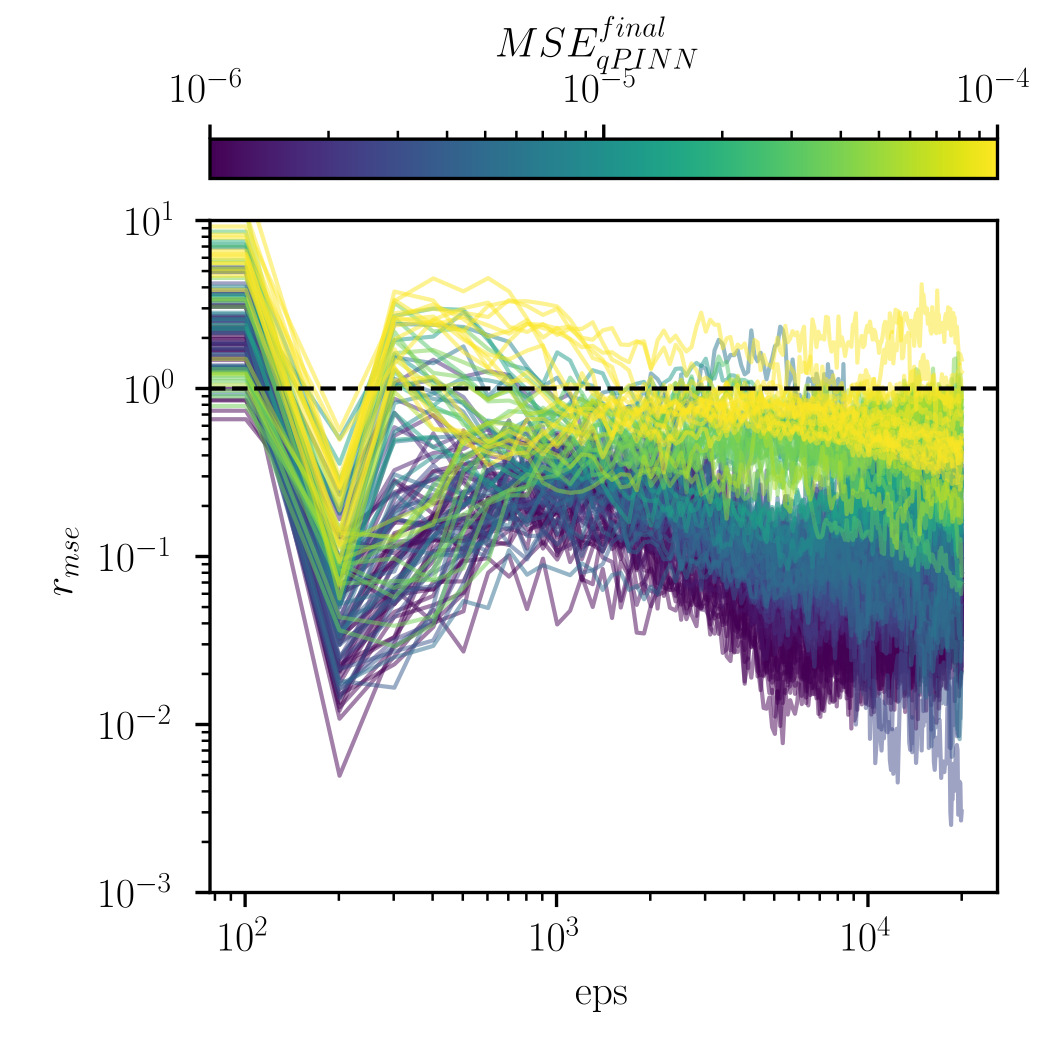}
        }
    \caption{MSE as a function of epoch ratios (top) and MSE ratios as a function of epochs trained (bottom) when training 'poly' boundary conditions and forcings. The line colors indicate the final MSE of the qPINN for the corresponding PDE. The dashed lines mark, where cPINNs and qPINNs perform equally.}
         \label{fig:poly_results}
\end{figure}

Similar to Fig.~\ref{fig:convergence_rate_ratio} and Fig.~\ref{fig:mse_ratio_xsin} for 'xsin', Fig.~\ref{fig:poly_results} visualizes the training progress of cPINNs and qPINNs for 'poly' boundary conditions and forcings. Again, the general accuracy limit has a strong influence on epoch ratios and MSE ratios, which show similar behavior as for 'xsin'. This indicates that qPINNs can offer benefits in various use cases.

\subsection{Parameter path in loss landscape}
\begin{figure*}[t]
    \centering
        \resizebox{0.85\textwidth}{!}
        {\includegraphics[width=\linewidth]{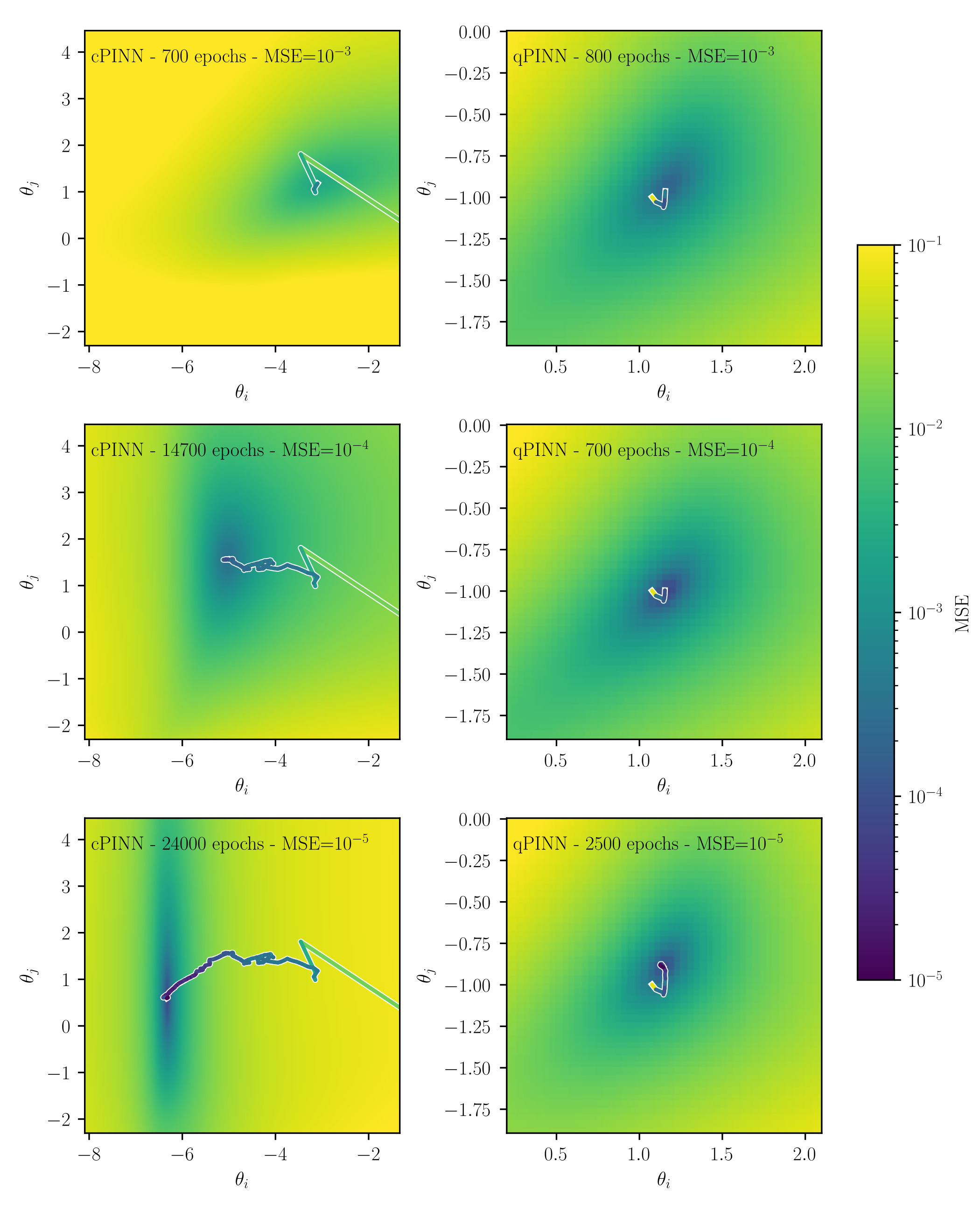}
        }
    \caption{The evaluation of two trainable parameters during training is illustrated when reaching a MSE of $10^{-3}$, $10^{-4}$, and $10^{-5}$ for a cPINN (left) respectively a qPINN (right). The lines track the parameter values from the initial value up to the visualized epoch. While for the cPINN the two parameters are weights from neighboring nodes from the first hidden layer, for the qPINN the parameters are two neighboring parameters from the first variational layer of the quantum circuit. The background image is obtained by calculating the MSE when varying the parameters $\boldsymbol{\theta}_i$ and $\boldsymbol{\theta}_j$ in the respective epoch. The total number of trainable parameters in the circuit is 250 and the amount of training data is 1024 for each, boundary and PDE loss. The considered PDE uses $L=0.1, N=1.0$ with 'xsin' boundary conditions and forcing.}
         \label{fig:path_loss_landscape}
\end{figure*}

To get an impression of the reason for the different convergence rates of cPINNs and qPINNs the evolution of randomly chosen parameter pairs are traced during training and visualized in Fig.~\ref{fig:path_loss_landscape}. It is clearly visible that the path of the cPINN's parameter pair is more chaotic while the qPINN's parameter pair moves much more targeted in the direction of final parameter values, making it more efficient. This seems to be made possible by a landscape that has a clearly recognizable and fixed minimum. This could potentially be caused by the entanglement in the quantum circuit, which connects the trainable parameters more strongly to each other than trainable parameters are connected in a classical network. Consequently, the impact on the landscape of other parameters when updating a given parameter can be minimized, resulting in a more stable and well-behaved optimization landscape for each of the parameters and therefore faster convergence in terms of training epochs. This is visible in Fig.~\ref{fig:path_loss_landscape}, where the loss landscape of the cPINN changes during the training, while the qPINN's loss landscape tends to remain constant. Although further research is required to validate the underlying explanation, the ability to efficiently identify a trajectory toward the optimal parameter configuration becomes increasingly critical as the considered problems and, consequently, the associated loss landscapes grow in complexity. This observation further suggests that qPINNs represent a promising approach for addressing such complex applications with improved efficiency.
\vfill

\end{document}